\documentclass{article}
\usepackage{spconf,amsmath,graphicx,hyperref,multirow,booktabs}
\usepackage{makecell}
\usepackage{cite}

\title{MSRBench: A Benchmarking Dataset for Music Source Restoration}

\name{
  \begin{tabular}{c}
  Yongyi Zang$^{1\star}$\thanks{$^*$ These authors contributed equally.} \qquad
  Jiarui Hai$^{2\star}$ \qquad
  Wanying Ge$^{1}$ \qquad
  Qiuqiang Kong$^{3}$ \\
  Zheqi Dai$^{3}$ \qquad
  Helin Wang$^{2}$ \qquad
  Yuki Mitsufuji$^{4}$ \qquad
  Mark D.\ Plumbley$^{5}$
  \end{tabular}
}

\address{
    $^1$Independent Researcher \qquad
    $^2$John Hopkins University \qquad \\
    $^3$The Chinese University of Hong Kong \qquad
    $^4$Sony AI \qquad
    $^5$King's College London
}

\begin{document}
\ninept
\maketitle

\begin{abstract}
Music Source Restoration (MSR) extends source separation to realistic settings where signals undergo production effects (equalization, compression, reverb) and real-world degradations, with the goal of recovering the original unprocessed sources. 
Existing benchmarks cannot measure restoration fidelity: synthetic datasets use unprocessed stems but unrealistic mixtures, while real production datasets provide only already-processed stems without clean references. We present \textbf{MSRBench}, the first benchmark explicitly designed for MSR evaluation. 
MSRBench contains raw stem-mixture pairs across eight instrument classes, where mixtures are produced by professional mixing engineers. These raw-processed pairs enable direct evaluation of both separation accuracy and restoration fidelity. Beyond controlled studio conditions, the mixtures are augmented with twelve real-world degradations spanning analog artifacts, acoustic environments, and lossy codecs. Baseline experiments with U-Net and BSRNN achieve SI-SNR of -37.8 dB and -23.4 dB respectively, with perceptual quality (FAD$_\text{CLAP}$) around 0.7-0.8, demonstrating substantial room for improvement and the need for restoration-specific architectures.
\end{abstract}
\begin{keywords}
Music Source Restoration, Benchmarking, Audio Restoration, Audio Signal Processing
\end{keywords}

\section{Introduction}
\label{sec:intro}

Music Source Restoration (MSR)~\cite{zang2025music} extends traditional Music Source Separation (MSS)~\cite{cano2018musical} to address the complexities of real-world audio production. While MSS assumes musical mixtures are linear combinations of clean sources, MSR recognizes that practical recordings undergo substantial transformations during production and transmission: equalization, dynamic range compression, harmonic distortion, reverberation, lossy codec artifacts, and acoustic degradations. MSR aims to recover the original, unprocessed sources from these degraded mixtures.

The MSS community has established rigorous evaluation frameworks through competitions like the Music Demixing Challenge (MDX)~\cite{mitsufuji2022music} and the Cinematic Sound Demixing tracks (CDX)~\cite{uhlich2023sound} at the Sound Demixing Challenge~\cite{fabbro2023sound}, where standard academic benchmarking datasets such as MUSDB18-HQ~\cite{rafii2019musdb18} and MoisesDB~\cite{pereira2023moisesdb} are utilized. Community benchmarks, such as the Multisong Dataset~\footnote{\url{https://mvsep.com/en/quality_checker}} also facilitates multi-faceted evaluation. These benchmarks provide high-quality stem-mixture pairs enabling standardized model comparison. However, they are insufficient for MSR evaluation: their ``ground truth'' stems already contain production effects (compression, reverb, EQ), making it impossible to measure restoration to truly unprocessed sources. Additionally, they do not evaluate models under real-world degradations like analog artifacts, acoustic environments, or codec compression; all of which are critical for MSR systems deployed in practice. Recent restoration methods such as VoiceFixer~\cite{liu2021voicefixer}, Apollo~\cite{li2025apollo}, and SonicMaster~\cite{melechovsky2025sonicmaster} demonstrate progress on specific degradation types, but without benchmarks providing unprocessed ground truth under diverse conditions, these approaches cannot be rigorously evaluated or compared.

The core challenge is data acquisition. While clean stems can be linearly summed to create synthetic mixtures for MSS, MSR requires aligned pairs of unprocessed stems and professionally mixed outputs, a resource rarely available at scale. Engineers' exported stems typically already include effects, while publicly available raw stems rarely have corresponding processed versions. This scarcity forces researchers into suboptimal strategies: applying additional degradations to already-processed datasets~\cite{melechovsky2025sonicmaster,li2025apollo,liu2021voicefixer}, which undermines the restoration objective; or simulating mixing effects on raw datasets~\cite{zang2025music}, which produces unrealistic mixtures and limits cross-study comparability. The lack of standardized evaluation hinders progress, obscuring genuine advances and preventing robust baseline establishment.

We introduce MSRBench, the first standardized benchmark for music source restoration with truly unprocessed ground truth. Developed in collaboration with professional mixing engineers, MSRBench provides 2,000 clips across eight instrument classes, each with parallel unprocessed stems and professionally mixed outputs. We augment these with 12 real-world degradations spanning analog artifacts, acoustic environments, and both traditional and neural codecs. To establish baselines, we adapt established MSS models, U-Net~\cite{stoller2018wave} and BSRNN~\cite{yu2022high} to the mel-band domain~\cite{wang2023mel} and train on the RawStems dataset~\cite{zang2025music}. Evaluation combines reference-based metrics (SI-SNR) measuring reconstruction fidelity with perceptual metrics (FAD$_\text{CLAP}$~\cite{wu2023large}) capturing output naturalness.

Beyond establishing baseline performance, MSRBench reveals fundamental challenges in music source restoration. While BSRNN achieves reasonable perceptual quality (FAD = 0.74), SI-SNR remains poor (-23.4 dB), primarily due to phase estimation errors. We observe that magnitude spectrograms align well with targets, and hypothesize phase misalignment may have severely degraded signal-level objective metrics. Instrument-level analysis shows restoration difficulty varies dramatically: bass achieves -17.3 dB while percussion fails catastrophically at -55.2 dB, revealing that harmonic content and temporal structure pose distinct restoration challenges. Degradation sensitivity differs by architecture: BSRNN handles DSP codecs (-23.6 dB) far better than analog artifacts (-28.6 dB). These findings indicate that restoration cannot be solved by simply adding robustness to separation models; MSR requires phase-aware architectures, instrument-specific strategies, and evaluation methods beyond traditional separation metrics.

We release MSRBench v1.1 at \url{https://huggingface.co/datasets/yongyizang/MSRBench/tree/v1.1}\footnote{This v2 corrects Table~\ref{tab:main-result} after the MSRBench v1.1 dataset fix: in v1, DT2/DT3 mixtures were identical to DT0, and DT11/DT12 partly duplicated each other. Results use dataset revision v1.1.} under CC-BY-NC license as well as evaluation protocols and baseline implementations under MIT License at \url{https://github.com/yongyizang/MSRKit} to foster reproducible research in this emerging field.

\section{The MSRBench Dataset}

MSRBench contains 2,000 professionally mixed 10-second clips at 48 kHz stereo: 250 clips for each of eight instrument classes (vocals, guitar, keyboard, synthesizer, bass, drums, percussion, orchestra). Each clip is evaluated under 13 conditions: the original mastered mixture (DT0) plus 12 degradation types, yielding 26,000 total stem-mixture pairs.

\subsection{Mixing Process}
Professional audio engineers produced MSRBench mixtures using industry-standard DAWs, preserving parallel tracks of \textbf{processed stems} (with effects) and \textbf{unprocessed stems} (dry references). The workflow follows standard production practice across three stages:

\textbf{Stem Processing.} Engineers applied per-instrument processing including equalization (high-pass filtering, resonance correction, tonal shaping), dynamic range compression, harmonic saturation, and spatial effects (reverb, delay, stereo panning).

\textbf{Mixing.} Processed stems are combined with gain-staging and bus processing (group compression, corrective EQ, shared reverb/delay) to establish acoustic coherence.

\textbf{Mastering.} Final processing includes broadband and multiband EQ, multiband compression, limiting for streaming platforms, and stereo width adjustment. The resulting \textbf{mastered mixture} serves as the reference condition (DT0) for all evaluation scenarios.

\subsection{Mixture Degradations}

To reflect realistic consumption and transmission conditions, MSRBench evaluates restoration under 12 degradations spanning three categories: analog/acoustic effects (DT1-DT4), traditional codecs (DT5-DT8), and neural codecs (DT9-DT12). The unmodified mastered mixture serves as the reference condition (DT0).

Analog and acoustic degradations simulate historical playback media and environmental capture. Radio (DT1) models FM broadcast reception using GNU Radio~\cite{blossom2004gnu} with Rayleigh fading~\cite{sklar2002rayleigh}, multipath propagation, and additive noise at 26 dB SNR. Cassette (DT2) uses our open DSP chain: pre-emphasis, saturation, head-bump EQ, high-frequency roll-off, wow/flutter, azimuth error, dropouts, and hiss. Vinyl (DT3) uses bandwidth limiting, cartridge resonance, crosstalk, groove-wear distortion, $33\frac{1}{3}$ rpm wow, rumble, surface noise, crackle, and pops.\footnote{Both chains: \url{https://huggingface.co/datasets/yongyizang/MSRBench/blob/v1.1/scripts/degrade.py}} Live sound (DT4) combines room acoustics (PyRoomAcoustics impulse responses~\cite{scheibler2018pyroomacoustics}), phone microphone bandpass filtering, and ambient noise from WHAM!~\cite{wichern2019wham}\footnote{\url{http://wham.whisper.ai/}} at 20 dB SNR, simulating concert recordings on mobile devices.

Traditional lossy codecs represent common streaming and storage scenarios: 64 kbps and 128 kbps for both AAC (DT5, DT7) and MP3 (DT6, DT8). These bitrates cover low-bandwidth streaming and typical portable device quality~\cite{tortosa2014optimal}.

Neural codecs capture artifacts from emerging compression methods. We apply Descript Audio Codec (DAC)~\cite{kumar2023high} at 22 kHz (DT9) and 44 kHz (DT10), and the official Encodec~\cite{defossezhigh} 48 kHz model with 6 kbps (DT11) and 3 kbps (DT12). Unlike traditional codecs, neural methods introduce distortions from discrete latent quantization rather than frequency-domain truncation, presenting distinct restoration challenges.

Together, these conditions span analog imperfections, environmental variability, and both legacy and modern compression artifacts, ensuring evaluation reflects the heterogeneity of real-world audio degradation.

\section{Experiments}
\begin{figure}[t]
  \centering
\includegraphics[width=0.4\textwidth]{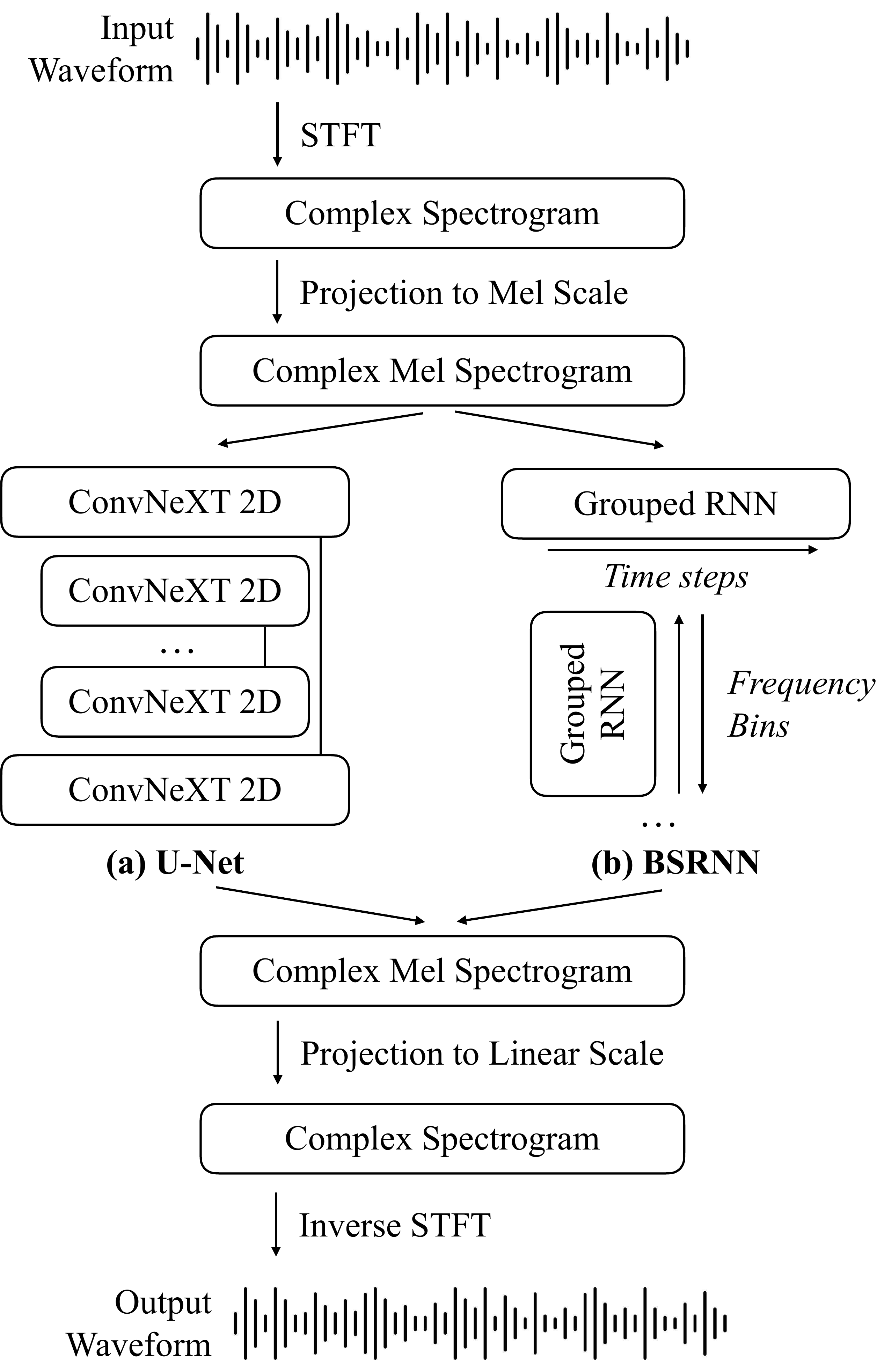}
  \caption{Overview of two baseline architectures. Both networks process complex mel spectrograms: The U-Net employs cascading upsampling and downsampling layers with skip connections, while the BSRNN uses grouped RNN networks operating in parallel to alternately model the temporal and frequency axes. The frequency-axis modeling is bidirectional, whereas temporal modeling is unidirectional (causal).}
  \label{fig:baseline}
\end{figure}

\subsection{Baseline Models}
To establish baseline performance on MSRBench, we train two models adapted from the music source separation literature: U-Net~\cite{stoller2018wave} and BSRNN~\cite{yu2022high}, both modified to operate in the mel-band domain~\cite{wang2023mel} following current best practices. In addition to standalone regression-based objectives, we incorporate adversarial training with the multi-frequency discriminator from EnCodec~\cite{defossezhigh}.

\subsection{Metrics}
We use two representative metrics in our evaluation:

\textbf{SI-SNR} (Scale-Invariant Signal-to-Noise Ratio) is a reference-based objective measure that quantifies the ratio of target signal power to residual noise after optimal scaling. It reflects reconstruction fidelity by indicating how closely the separated waveforms match the ground-truth sources.

\textbf{FAD\textsubscript{CLAP}} is a perceptual measurement that computes the Fréchet distance between distributions of reference and separated signals embedded in the CLAP embedding space~\cite{wu2023large}. It reflects perceptual plausibility by capturing how natural and realistic the outputs sound to human listeners.

\subsection{Training Objective}
We adopt a multi-component loss function that balances reconstruction fidelity with perceptual quality. The total loss combines three complementary objectives:

\textbf{Reconstruction Loss.} To ensure accurate signal recovery, we employ a multi-scale mel-spectrogram reconstruction loss that measures the L1 distance between the predicted and ground-truth sources across multiple temporal resolutions. This is computed using three STFT configurations with window sizes of 512, 1024, and 2048 samples, weighted by $\lambda_{\text{recon}} = 45.0$. This multi-scale approach captures both fine-grained spectral details and broader temporal structure.

\textbf{Adversarial Loss.} To improve perceptual quality, we incorporate adversarial training using a hinge loss formulation, weighted by $\lambda_{\text{gan}} = 0.1$. The discriminator is trained to distinguish between real ground-truth stems and generated outputs, encouraging the generator to produce more natural-sounding separations that align with the distribution of real audio.

\textbf{Feature Matching Loss.} To stabilize adversarial training and provide additional perceptual guidance, we include a feature matching loss that minimizes the L1 distance between discriminator activations for real and generated samples, weighted by $\lambda_{\text{feat}} = 10.0$. This encourages the generator to match intermediate representations rather than directly fooling the discriminator, leading to more robust training.

The combined objective is:
\begin{equation}
\mathcal{L} = \lambda_{\text{recon}}\mathcal{L}_{\text{recon}} + \lambda_{\text{gan}}\mathcal{L}_{\text{gan}} + \lambda_{\text{feat}}\mathcal{L}_{\text{feat}}
\end{equation}

This formulation has proven effective for audio generation tasks and is particularly well-suited to MSR, where both reconstruction accuracy and perceptual naturalness are critical.

\subsection{Training Configuration}
Baseline models are trained using the following settings:

\textbf{Training Data.} We train all models on 4-second 48 kHz clips randomly sampled from stems in the RawStems dataset~\cite{zang2025music}. Baseline models use either convolution or recurrent networks to model the temporal axis, which scales naturally over time, therefore avoiding train/test mismatch. Data augmentation is applied by mixing target stem with other stems at SNR levels between 0 and 10 dB, while excluding clips below $-40$ dB RMS to avoid near-silent regions~\footnote{More details on training configurations can be accessed at \url{https://github.com/yongyizang/MSRKit}}.

\textbf{Architectures.} Both baselines are adapted to operate in the mel-band domain~\cite{wang2023mel}. For adversarial training, the multi-frequency discriminator is configured with three window sizes (2048, 1024, and 512 samples) and applies layer normalization. These design choices enable the baselines to capture both fine-grained spectral structure and perceptual quality.

\textbf{Optimization.} Models are optimized with Adam using a learning rate of $2 \times 10^{-4}$, momentum parameters $\beta_1 = 0.8$, $\beta_2 = 0.99$, and a 5{,}000-step warm-up schedule. Training is performed with batch sizes of 16 for U-Net and 8 for BSRNN, using 32 data loader workers. Each run spans 1 million steps (500k generator and 500k discriminator updates), requiring approximately 5 days on dual NVIDIA RTX 4090 GPUs.

\textbf{Inference and Evaluation.} During inference, we process each 10-second sample from MSRBench individually through the model without additional windowing, unlike common evaluation protocols in music source separation challenges~\cite{mitsufuji2022music, solovyev2023benchmarks, fabbro2023sound}. We aggregate results hierarchically: we first compute metrics per degradation type, then averaging across degradation types within each major category to produce category-level scores.

\begin{table*}[h]
\centering
\small
\caption{Results on MSRBench v1.1 under four evaluation settings: Original mixtures, Analog degradations, DSP-based degradations, and Neural codec degradations. Metrics reported are SI-SNR (dB) and FAD$_\text{CLAP}$. Instrument abbreviations: Gtr. = Guitar, Key. = Keyboard, Perc. = Percussion, Orch. = Orchestra.}
\begin{tabular}{cccccccccccc}
\toprule
\textbf{Setting} & \textbf{Model} & \textbf{Metric}
  & \textbf{Vocals} & \textbf{Gtr.} & \textbf{Key.} & \textbf{Synth}
  & \textbf{Bass} & \textbf{Drums} & \textbf{Perc.} & \textbf{Orch.} & \textbf{Averaged} \\
\midrule
\multirow{4}{*}{\makecell[c]{\textbf{Original}\\\textbf{(DT0)}}}
  & \multirow{2}{*}{U-Net}   & SI-SNR & -30.1 & -38.3 & -38.3 & -43.5 & -17.4 & -31.3 & -59.6 & -43.6 & -37.8 \\
  &                          & FAD    & 0.46 & 0.64 & 0.83 & 0.84 & 0.66 & 0.80 & 1.45 & 0.83 & 0.81 \\
  & \multirow{2}{*}{BSRNN}   & SI-SNR & -10.5 & -19.4 & -22.0 & -23.0 & -17.3 & -11.7 & -55.2 & -28.1 & -23.4 \\
  &                          & FAD    & 0.32 & 0.37 & 1.07 & 0.89 & 0.70 & 0.61 & 1.19 & 0.76 & 0.74 \\
\midrule
\multirow{4}{*}{\makecell[c]{\textbf{Analog}\\\textbf{(DT1--DT4)}}}
  & \multirow{2}{*}{U-Net}   & SI-SNR & -30.7 & -38.2 & -38.2 & -43.1 & -21.2 & -34.3 & -59.3 & -43.7 & -38.6 \\
  &                          & FAD    & 0.46 & 0.62 & 0.82 & 0.85 & 0.69 & 0.84 & 1.43 & 0.83 & 0.82 \\
  & \multirow{2}{*}{BSRNN}   & SI-SNR & -17.7 & -24.1 & -27.7 & -30.9 & -18.1 & -21.5 & -56.1 & -32.8 & -28.6 \\
  &                          & FAD    & 0.35 & 0.40 & 1.05 & 0.96 & 0.76 & 0.70 & 1.16 & 0.77 & 0.77 \\
\midrule
\multirow{4}{*}{\makecell[c]{\textbf{DSP}\\\textbf{(DT5--DT8)}}}
  & \multirow{2}{*}{U-Net}   & SI-SNR & -30.1 & -38.3 & -38.5 & -42.9 & -17.4 & -31.4 & -58.0 & -43.3 & -37.5 \\
  &                          & FAD    & 0.47 & 0.63 & 0.83 & 0.85 & 0.66 & 0.79 & 1.42 & 0.82 & 0.81 \\
  & \multirow{2}{*}{BSRNN}   & SI-SNR & -10.5 & -19.6 & -22.1 & -23.3 & -17.4 & -11.9 & -55.9 & -28.3 & -23.6 \\
  &                          & FAD    & 0.32 & 0.38 & 1.06 & 0.90 & 0.71 & 0.62 & 1.21 & 0.75 & 0.74 \\
\midrule
\multirow{4}{*}{\makecell[c]{\textbf{Neural Codec}\\\textbf{(DT9--DT12)}}}
  & \multirow{2}{*}{U-Net}   & SI-SNR & -30.9 & -38.8 & -37.8 & -43.5 & -17.4 & -31.8 & -59.3 & -43.1 & -37.8 \\
  &                          & FAD    & 0.50 & 0.73 & 0.90 & 0.88 & 0.73 & 0.85 & 1.52 & 0.86 & 0.87 \\
  & \multirow{2}{*}{BSRNN}   & SI-SNR & -9.9 & -21.7 & -25.3 & -27.6 & -17.3 & -14.2 & -60.6 & -30.7 & -25.9 \\
  &                          & FAD    & 0.43 & 0.54 & 1.15 & 0.97 & 0.81 & 0.72 & 1.28 & 0.80 & 0.84 \\
\bottomrule
\end{tabular}
\label{tab:main-result}
\end{table*}

\subsection{Results}

Table~\ref{tab:main-result} presents baseline performance across all evaluation conditions. BSRNN substantially outperforms U-Net on SI-SNR, achieving -23.4 dB versus -37.8 dB on original mixtures (DT0), while FAD$_\text{CLAP}$ scores remain comparable at 0.74 and 0.81 respectively. This discrepancy between reconstruction accuracy and perceptual quality persists across all degradation types, despite their strong performance on source separation tasks\cite{stoller2018wave, jansson2017singing, kadandale2020multi, luo2023music, lu2024music, luo2024improving}, suggesting MSR requires different approaches.

Degradation effects reveal distinct model characteristics. U-Net maintains stable SI-SNR between -38.6 and -37.5 dB across the degradation categories in Table~1. BSRNN shows greater sensitivity: analog effects cause the largest performance drop (-28.6 dB), DSP codecs have minimal impact (-23.6 dB), and neural codecs fall between these extremes (-25.9 dB). BSRNN's recurrent architecture appears to handle codec artifacts more effectively than analog and acoustic distortions.

Instrument-level analysis shows substantial variation in restoration difficulty. For BSRNN, bass achieves -18.1 to -17.3 dB SI-SNR while percussion fails catastrophically at -60.6 to -55.2 dB. FAD$_\text{CLAP}$ rankings differ: vocals achieve 0.32-0.43 while percussion (1.16-1.28) and keyboards (1.05-1.15) score substantially higher. Transient-rich sources fail on both metrics, while harmonically complex sources show better reconstruction accuracy but poorer perceptual quality.

\begin{figure*}[h!]
  \centering
\includegraphics[width=\textwidth]{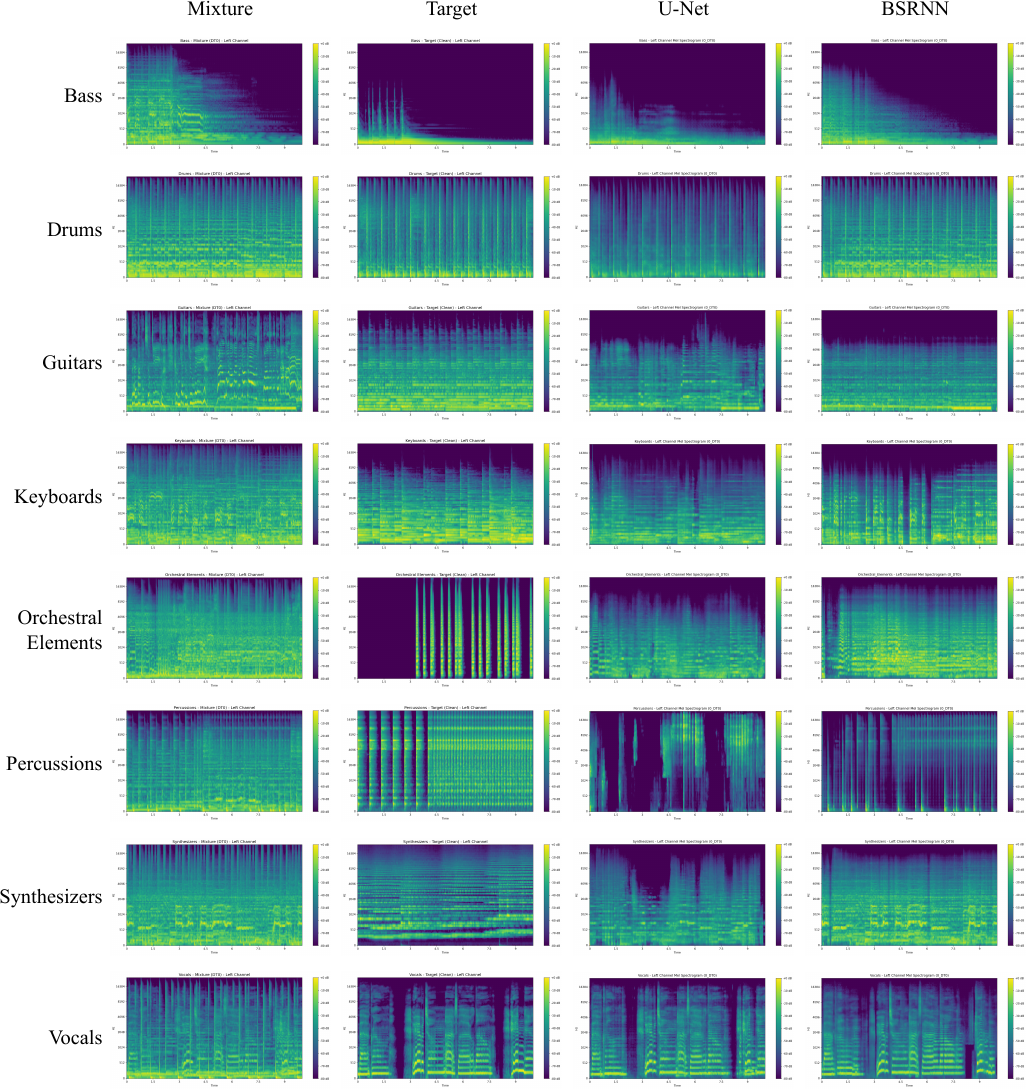}
\caption{Mel spectrogram examples showing mixture, target, and predictions from both baseline models. Despite poor SI-SNR values, predictions exhibit mostly correct spectral structure and temporal evolution, with primary artifacts appearing as reduced fine-grained detail, incomplete instrumental removal rather than catastrophic distortion. Notably, vocal estimation is almost completely correct, yet scores very low SI-SNR.}
  \label{fig:examples}
\end{figure*}

The gap between SI-SNR and FAD$_\text{CLAP}$ warrants closer examination. Figure~\ref{fig:examples} shows mel spectrograms for both models alongside mixture and target stems. Despite poor SI-SNR values, separated sources exhibit largely correct spectral envelopes and temporal structure. Listening tests confirm this observation: outputs sound substantially better than -23 dB SI-SNR would suggest. Magnitude spectrograms align well with targets, indicating that phase misalignment is the primary source of degradation in objective metrics. Standard evaluation protocols apply 512-sample sliding windows which partially address this issue, but we evaluate on full 10-second segments as accurate phase estimation is fundamental to restoration quality.

Current objective metrics do not fully capture restoration performance. The results suggest that future work requires either subjective evaluations or reference-free perceptual metrics~\cite{chinen2020visqol, tjandra2025meta, alakuijala2025zimtohrli} that better align with human judgment of restoration quality.

\section{Discussion}

Source separation and source restoration exist on a continuum between recovery and generation. Traditional source separation assumes clean sources and solves the recovery problem by imposing constraints on the separated components: statistical independence~\cite{hyvarinen2000independent}, sparsity~\cite{vincent2006performance}, or non-negativity~\cite{fevotte2009nonnegative}. Data-driven approaches replace these analytical constraints with learned priors from training data~\cite{stoller2018wave}. However, the information-theoretic reality remains: lost information cannot be recovered without reconstruction through learned generative models.

Music source restoration extends further into generative territory. Production effects and degradations destroy information irreversibly: compression discards frequency components, reverb obscures temporal boundaries, codec quantization loses spectral detail. As degradation severity increases, restoration becomes increasingly speculative. Models must generate plausible content through learned priors rather than recover known signals. This explains the SI-SNR/FAD discrepancy in our results: SI-SNR penalizes any deviation from ground truth, while FAD measures distributional plausibility. Models can generate perceptually realistic outputs that differ from reference signals yet fulfill their intended purpose.

This pattern appears across domains where generation supplants reconstruction. Language model priors enable automatic speech recognition to achieve semantic accuracy despite word-level errors~\cite{chelba2012large, radford2023robust, xu2025qwen3}. Text-only language models answer visual~\cite{chen2024we} and music~\cite{zang2025you} domain questions through learned cross-modal associations. Generative models perform language-queried separation without task-specific training~\cite{lee2025dgmo, huang2025zerosep}. In each case, traditional reconstruction metrics underestimate system capability: word error rate, signal-to-noise ratio, and mean-squared error assume unique correct answers, while generative tasks admit multiple valid solutions.

Our results demonstrate this evaluation gap concretely. Models achieve reasonable perceptual quality (FAD = 0.74) while failing reconstruction metrics (SI-SNR = -23.4 dB). Since humans are the final users of these systems, evaluation should prioritize perceptual alignment and task utility over exact reconstruction. As generative approaches become standard, benchmarks must evolve to measure what matters: whether outputs serve their intended use, not whether they match ground truth precisely.

\section{Conclusions}
We introduce MSRBench, the first music source restoration benchmark with unprocessed ground truth, containing 2,000 professionally mixed clips across eight instruments and 13 degradation conditions. Baseline experiments reveal that current separation architectures are insufficient: BSRNN achieves reasonable perceptual quality (FAD = 0.74) but poor reconstruction (SI-SNR = -23.4 dB). Qualitative analysis identifies phase estimation as the primary bottleneck—magnitude spectra align well, but phase errors severely degrade objective metrics. Restoration difficulty varies dramatically by source: harmonic instruments restore substantially better than transient-rich percussion, indicating fundamental gaps in temporal modeling. These findings point to three research directions: phase-aware architectures, instrument-specific strategies, and better evaluation metrics. MSRBench v1.1 is available at \url{https://huggingface.co/datasets/yongyizang/MSRBench/tree/v1.1}; protocols and baselines are at \url{https://github.com/yongyizang/MSRKit}.

\section*{Acknowledgments}
We thank Youran Ni (Wuhan University) for reporting the dataset issue.

\newpage
\bibliographystyle{IEEEbib}
\bibliography{strings,refs}

\end{document}